\documentclass[aip,reprint]{revtex4-1}
\usepackage{latexsym, amssymb, mathrsfs, graphicx}

\begin{document}


\title{Multi-chord fiber-coupled interferometer with a long coherence length laser} 



\author{Elizabeth C. Merritt}
\affiliation{Physics and Astronomy, University of New Mexico, Albuquerque, NM, 87131, USA}
\author{Alan G. Lynn}
\affiliation{Electrical and Computer Engineering, University of New Mexico, Albuquerque, NM, 87131, USA}
\author{Mark A. Gilmore}
\email{gilmore@ece.unm.edu}
\affiliation{Physics and Astronomy, University of New Mexico, Albuquerque, NM, 87131, USA}
\affiliation{Electrical and Computer Engineering, University of New Mexico, Albuquerque, NM, 87131, USA}
\author{Scott C. Hsu}
\affiliation{Physics Division, Los Alamos National Laboratory, Los Alamos, NM, 87545, USA}

\date{\today}

\begin{abstract}
 This paper describes a 561 nm laser heterodyne interferometer that provides time-resolved measurements of line-integrated plasma electron density within the range of $10^{15}-10^{18} \mbox{ cm}^{-2}$. Such plasmas are produced by railguns on the Plasma Liner Experiment (PLX), which aims to produce $\mu$s-, cm-, and Mbar-scale plasmas through the merging of thirty plasma jets in a spherically convergent geometry. A long coherence length, 320 mW laser allows for a strong, sub-fringe phase-shift signal without the need for closely-matched probe and reference path lengths. Thus only one reference path is required for all eight probe paths, and an individual probe chord can be altered without altering the reference or other probe path lengths. Fiber-optic decoupling of the probe chord optics on the vacuum chamber from the rest of the system allows the probe paths to be easily altered to focus on different spatial regions of the plasma. We demonstrate that sub-fringe resolution capability allows the interferometer to operate down to line-integrated densities of order $10^{15} \mbox{ cm}^{-2}$.
\end{abstract}


\maketitle 

\section{Introduction}
The study of merging supersonic plasma jets is rich in potential physics, as well as serving as a potential method of forming imploding plasma liners relevant to fundamental scientific studies of Mbar-scale, high-energy-density (HED) plasmas and potentially magneto-inertial fusion (MIF). \cite{Hsu} Determining plasma electron density, $n_e$, during jet propagation and jet merging is imperative for understanding the underlying plasma dynamics. Electron density dynamics can yield information on jet expansion, potential shock structure formation, and kinetic energy changes across any merging boundaries. The Plasma Liner Experiment \cite{PLX} aims to study the merging dynamics of multiple plasma jets with the ultimate goal of creating a spherical plasma liner through the merging of thirty jets in spherically convergent geometry. To this end, an 8 chord, visible, heterodyne, fiber-optic based interferometer has been designed and built for measuring $n_e$ during experiments on one-jet propagation and two-jet merging, as well as the merging of thirty plasma jets. Individual plasma jets, which are generated and accelerated by railguns manufactured by HyperV Technologies, \cite{rsi-witherspoon, aps-witherspoon} are expected to have initial density $n_e \approx 10^{15} - 10^{17} \mbox{ cm}^{-3}$ , temperature 1-5 eV, and velocity $\approx 50$ km/s, corresponding to Mach numbers of $M \approx 10-35$.

The visible wavelength of the interferometer was chosen to accommodate a density range from that of an individual jet with peak line-integrated densities of $\int n_e dl = 10^{16}-10^{17} \mbox{ cm}^{-2}$, through at least density doubling during the merging of two jets, and finally through $\int n_e dl = 10^{18} \mbox{ cm}^{-2}$ corresponding to part way through liner convergence. Fiber-optic coupling between the laser source and the probe path launch and collection optics allows for long probe path lengths and for the large optical table to be placed several meters away from the plasma chamber. The long coherence length of the laser ($>10$ m) allows path length variance between the probe and reference paths without the need to tailor the path length difference to the periodicity of the laser coherence length. \cite{Kumar} The long coherence length of the solid state laser, in conjunction with the fiber-optic setup, allows easy reconfiguration of the chord geometry. For example, the chord positions can be configured to measure the density profile of a single jet at different points along the axis of jet propagation, or for simultaneous measurements of the plasma density before and after a merging boundary between two jets.

\section{Equipment}

\subsection{Optics}
The fiber optic components of the interferometer split the optics into three functionally decoupled stages: the initial establishment of the probe and reference chords, the chord positioning optics on the chamber, and the final probe/reference beam recombination. The initial establishment of the chords and the final recombination take place on a 3.05 m x 1.22 m optical table located at a distance of approximately 6.1 m from the chamber. One of the primary advantages of fiber-coupling is the ability to place the bulk of the optics at a distance from the chamber without the need to preserve direct line-of-sight paths between the optics.

An Oxxius 561-300-COL-PP-LAS-01079, diode-pumped solid-state, 320 mW, 561 nm laser \cite{specs} produces the initial beam. A 50 percent attenuation filter is attached to the aperture of the laser, bringing the transmitted laser power down to 160 mW.  A beam divergence of 1.2 mrad, a path length of $\approx 2$ m before being coupled into the fiber, and an initial beam diameter of 0.6 mm renders collimation optics in this stage unnecessary. The coherence length of the Oxxius laser is at least 10m as conservatively specified by the manufacturer, \cite{specs} but the theoretical coherence length is much higher as estimated from the laser frequency linewidth \cite{specs} of 1 MHz as follows: \cite{Born}
\begin{eqnarray}
\Delta l = c \Delta t \approx \frac{c}{\Delta \nu} = \frac{c}{1 \mbox{ MHz}} \approx 100 \mbox{ m}.
\end{eqnarray}

As long as length differences between the reference and probe paths are kept small compared to the laser coherence length, fringe visibility will still be high. \cite{Born, Shajenko} Bench tests of one chord of the interferometer allowed at least a 2.4 m path length mismatch between the probe and reference chord without any appreciable signal degradation. This allows the use of one reference chord with respect to all eight probe chords, since the probe chord length variation due to the optics' placement on the chamber is easily kept below 2.4 m.

Thorlabs 460HP single-mode fiber is used for all fiber paths before reference/probe beam recombination. Single-mode fiber, while requiring greater laser power due to poorer coupling into the small fiber core, bypasses any signal degradation due to the interference effects from uneven wave-packet spreading and delay in different transverse lasing modes in multi-mode fiber. \cite{Shajenko, Hecht} Thus, single-mode fiber preserves the interferometer's ability to function with chord path length differences. Thorlabs BFH48-400 multi-mode fibers is used for carrying the final signal to the electronics since the phase relationship between the light of the two chords is established in the interference pattern and coherence concerns are no longer an issue.

\begin{figure*}
  \includegraphics{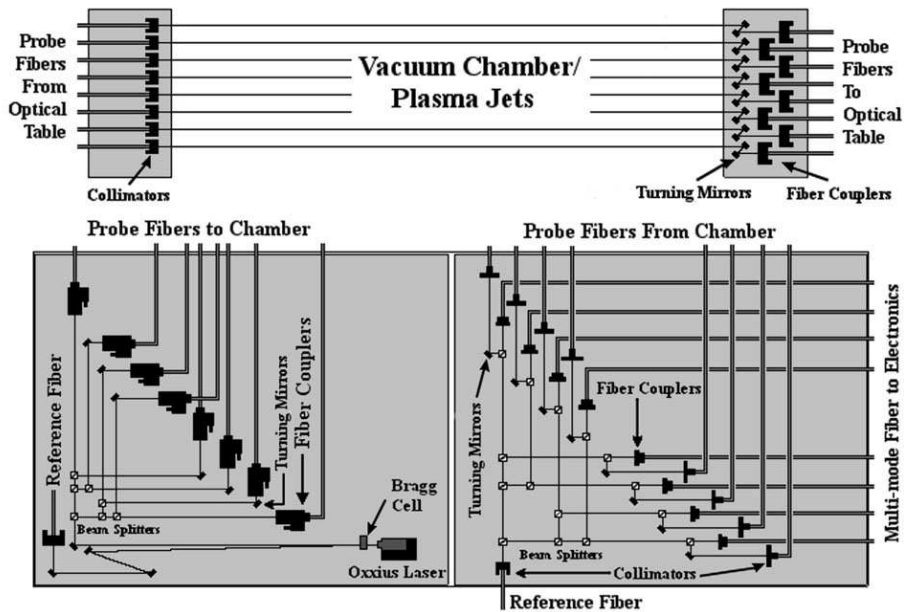}\\
  \caption{Interferometer optics layout, showing optical component layout for all three fiber-coupled stages: the establishment of the probe and reference chords, the chord-positioning optics on the chamber, and the final probe/reference beam recombination.}
\end{figure*}

An IntraAction ATM-1102DA1B Bragg cell beam splitter splits the original laser beam into the reference beam, now with a frequency shift of 110 MHz, and an unaltered beam, which is then split seven beam splitters into the eight probe beams. The Bragg cell power splitting is uneven, making the reference beam $\approx 20$ mW and the initial unmodulated beam 140 mW, which is filtered with a 0.3 OD neutral density filter. After splitting, all beams are attenuated to $<$ 8.3 mW and then routed to the first set of fiber-couplers, as shown in Fig 1. The reference beam is coupled, using a Thorlabs PAF-X-18-PC-A fiber coupler in a K6X kinematic mount, into a 42 m, single-mode fiber and routed directly to the recombination optics. Each probe chord is coupled using a Newport F-91-C1 fiber coupler into a 20 m, single-mode fiber. These fibers run from the optical table containing all previous optics to a 61 cm x 30.5 cm optical breadboard mounted on a 75 cm flange on the vacuum chamber. The breadboard is mounted with the 61 cm edge of the board adjacent to the 57 cm edge of a 57 cm x 19.8 cm x 2.54 cm borosilicate window.

The probe fibers terminate at a set of eight Thorlabs CFC-11X-A adjustable focal length collimators in Thorlabs KM100T kinematic mounts. Each collimator produces a collimated beam of $\approx 3$ mm diameter over a distance of 3 m. All eight probe beams pass through the chamber to an identical rectangular window/optical breadboard setup on a port on the opposite hemisphere of the 2.74 m diameter spherical chamber. A set of eight Thorlabs fiber-couplers are mounted on the second breadboard. Turning mirrors on either or both breadboards route the probe beams from the collimators to the fiber-couplers, where the beams are coupled into a second set of 20 m single-mode fibers. The exact optics arrangement varies depending on the physics being studied. Optics may be positioned to measure the plasma density through the center of a single jet at eight different points along the axis of jet propagation; this arrangement is optimal for studying electron density evolution in the jet as it propagates. The optics may also be positioned to measure the plasma density of the jet at one point along the axis of jet propagation but at different radial positions within the jet; this arrangement is optimal for a radial density profile measurement. The other major advantage of the fiber-optic system is the decoupling of the optics that dictate chord placement in the plasma from the rest of the system. This decoupling reduces the task of altering chord arrangement to the placement of approximately 32 components (four per chord) instead of the realignment of all 110 optical components in the system.

All probe and reference fibers are routed to the recombination optics. Each fiber terminates at an adjustable focus collimator, where they are again collimated into $\approx 3$ mm diameter beams. Beam splitters split the reference beam into eight beams, one for recombination with each probe chord. Each of the reference beams pass through another beam splitter, where the recombination with the probe beam occurs, and to the final Thorlabs fiber coupler, mounted in a simple CP02FP fiberport mount. Each probe beam exits the fiber and reflects off a turning mirror into the beam splitter containing the corresponding reference beam. At the final fiber couplers the light containing the interference signal is coupled into multi-mode fibers.

Since the beams have passed through different fibers, the polarization rotating effect of the fibers must be considered. Light emitted from the fibers is plane polarized, so to maximize the interference present when the probe and reference beams overlap, the polarizations of the light in both beams must be parallel to each other.\cite{Hecht, Born} Each probe chord collimator is held in a rotatable mount which allows each probe beam polarization to be independently matched to the reference beam polarization.

\subsection{IF Electronics}
 The Bragg cell (acousto-optic modulator, AOM) operation requires an IntraAction ME-1002 RF generator that produces a 110 MHz, 2 W signal. This signal passes through an 8.9 dB directional coupler, transmitting a small portion of the 110 MHz signal to the processing electronics while the rest of the signal is transmitted to the Bragg cell. The processing portion of the 110 MHz signal is transmitted to a Mini-Circuits ZCS-8-13-S+ eight-way electronic splitter, creating one 110 MHz Local Oscillator (LO) signal for each chord. Each LO signal is then transmitted to a Pulsar Microwave IDO-05-412 IQ Demodulator, as shown in Fig. 2.

The multi-mode fibers carrying the interference signals terminate at a set of Thorlabs PDA10A photoreceivers. The output signal from the photoreceivers pass through Lark Engineering MC110-55-6AA bandpass filters of 110 $\pm$ 55 MHz to filter out high-frequency components of the heterodyne-mixing, electromagnetic interference from the pulsed-power system, and most low-frequency electrical noise. The filtered signal is transmitted to the RF channel of an IQ Demodulator which decomposes the interference signal into two signals, I and Q, proportional to the sine and cosine of the frequency difference between the signal frequency and the LO frequency. The I and Q signals pass through a low-pass filter of maximum frequency 40 MHz, which removes any remaining high-frequency noise. Line-integrated electron densities of the plasma are extracted from the remaining frequencies contained in the I and Q signals.
\begin{figure*}
\begin{center}
  \includegraphics{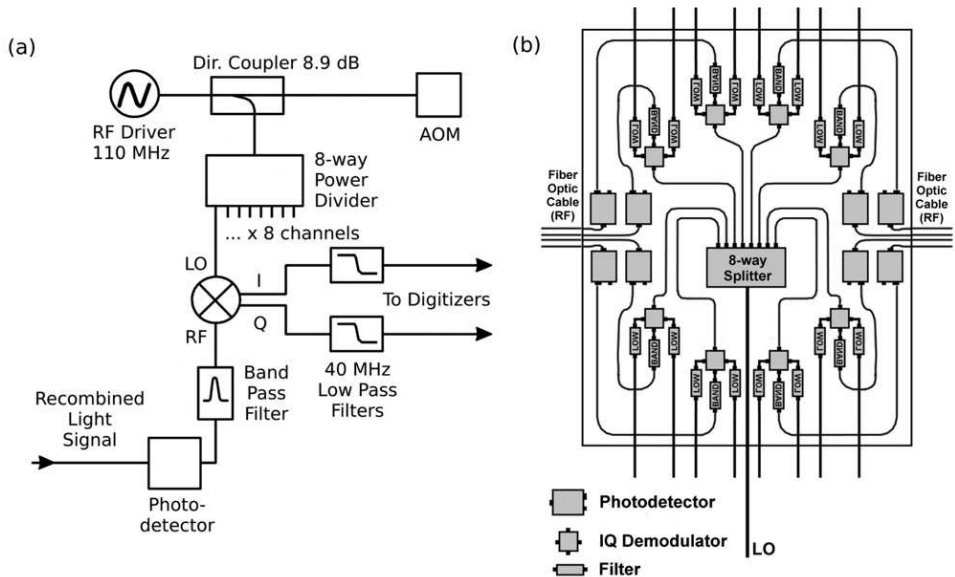}\\
  \caption{(a) An electrical schematic of RF components for all eight interferometer channels. (b) A layout of the RF electronics for all eight interferometer channels.}
  \end{center}
\end{figure*}

\section{Analysis of Beam Propagation and Light Background}
Visible interferometry was chosen by 2D ray-tracing simulation which included light deflection, fringe shift, and free-free Bremstrahlung absorption as the mechanism of attenuation. The laser chords were assumed to pass through a circularly symmetric plasma of radius 35 cm, with an exponentially-decaying radial profile. This plasma is a 2D approximation of the plasma liner expected to be generated by all 30 railguns. The plasma maximum and minimum densities were assumed to be $10^{19} \mbox{ cm}^{-3}$ and $10^{16} \mbox{ cm}^{-3}$ respectively. The simulation found that laser phase shift is the limiting factor for the interferometer wavelength selection. Line-integrated electron densities of $10^{17}- 10^{18} \mbox{ cm}^{-2}$ (corresponding to path lengths of 10-20 cm through the plasma, for the plasma region with densities between $10^{16}- 10^{17} \mbox{ cm}^{-3}$) generated a phase-shift in excess of $N=15$ fringes, where $N=15$ is an approximate upper limit on the time-resolvable number of fringes. At the phase shift limit, a 561 nm beam generated a laser deflection of $ < 0.5 \times 10^{-3}$ m and a laser intensity of $ > 95 \%$ of the original intensity. Thus, visible interferometry is suitable for density measurements of the outer regions of the full plasma liner.

Early tests of the Mark I plasma railguns, \cite{MarkII} performed at HyperV Technologies, indicated a plasma density of order $10^{17} \mbox{ cm}^{-3}$ and diameter 5 cm, which correspond to a line-integrated electron density of $\approx 10^{18} \mbox{ cm}^{-2}$ . These values yield the same order of line integrated plasma densities as the outer regions of the simulated plasma profile. Thus, visible interferometry should be viable for single- and several-jet density experiments as well as for the thirty-jet experiments. This is further confirmed by HyperV Technologies' use of a HeNe interferometer \cite{Case} for their single plasma jet experiments.

\section{Experimental Results}

\begin{figure*}
\begin{center}
  \includegraphics{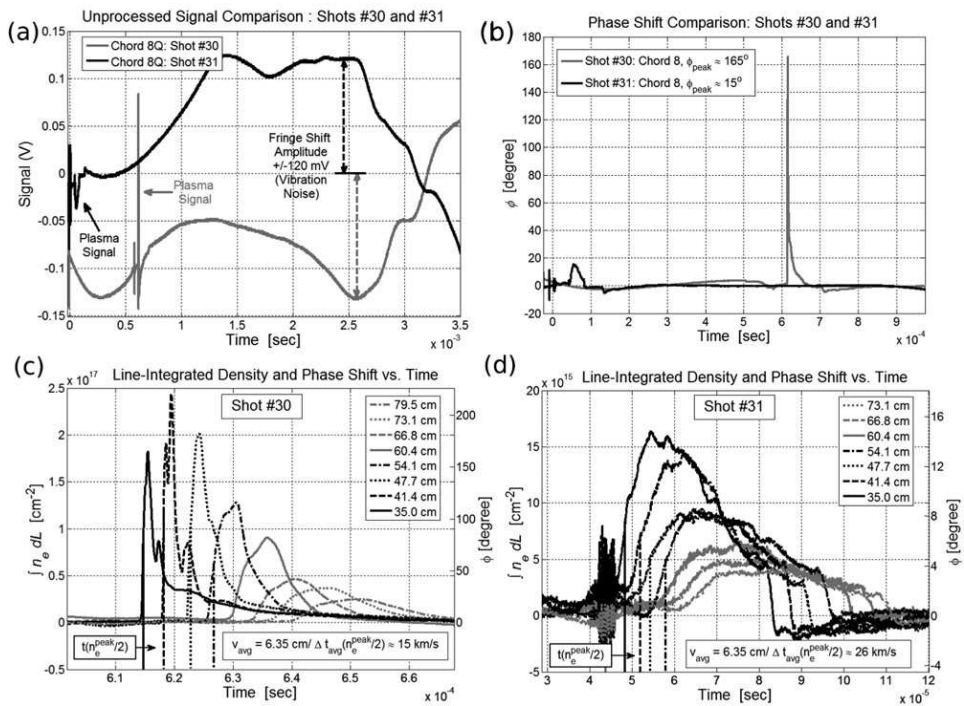}\\
  \caption{(a) A comparison of the raw signal amplitudes of Shots 30 and 31 for signal 8Q. (b) A comparison of the phase shift amplitudes (in degrees) after background noise has been subtracted between the same chord on Shots 30 and 31. (c) The line-integrated electron density (left axis) and phase shift (right axis) vs time for Shot 30 and (d) Shot 31 for all eight interferometer chords, where the chords are designated by their distances from the railgun muzzle. }
  \end{center}
\end{figure*}

PLX single-jet studies with the HyperV Mark I plasma railgun at $<300$ kA of peak current examine jet propagation at the lower end of the plasma density range detectable by this interferometer, $\int n_e  dl \approx 10^{15}-10^{17} \mbox{ cm}^{-2}$. Fig. 3 presents interferometry results from shot 30, typical of a ``higher" density shot, $\int n_e  dl \approx 10^{17} \mbox{ cm}^{-2}$, and shot 31, typical of a ``lower" density shot, $\int n_e  dl \approx 10^{16} \mbox{ cm}^{-2}$. Line-integrated electron densities are calculated as $\int n_e dl = 4 \pi \epsilon_0 m_e c^2/(e^2 \lambda_0) \Delta \phi$. As demonstrated in Fig. 3(a) and (b), the interferometer is capable of the sub-fringe measurements necessary for these densities. Fig 3(a) demonstrates the raw Q signals for Chord 8, for Shots 30  and 31 respectively. For Chord 8, the fringe amplitude of the electronic signal is approximately $+/-120$ mV, and the plasma signals, while only a fraction of the total fringe amplitude, are distinguishable from the bit noise as well as the lower-frequency vibrational noise. Figs. 3(b)-3(d) collectively show that a line-integrated electron density of $\int n_e dl \approx 1 \times 10^{17} \mbox{ cm}^{-2}$ corresponds to a phase-shift of $\Delta \phi \approx 90^o$, $\int n_e dl \approx 1 \times 10^{16} \mbox{ cm}^{-2}$ corresponds to $\Delta \phi \approx 9^o$, and $\int n_e dl \approx 5 \times 10^{15} \mbox{ cm}^{-2}$ corresponds to $\Delta \phi \approx 4^o$.

The average electronic signal per chord is approximately $+/-100$ mV with a bit noise amplitude of $+/-2$ mV. Since the raw signals show little noise on the time scale of the plasma interaction with the interferometer, the limiting signal noise is the bit noise from the digitizers. Thus, the interferometer can detect sub-fringe shifts of approximately $4^{o}$ with a signal to noise ratio of $2:1$ without signal processing. To further reduce the effects of bit noise, all signals are smoothed during processing using the MATLAB `filter' function to return a data array, $Signal_{new}$, where for every point in the array, $Signal_{new}(n) = \frac{1}{N} \Sigma_{i = n-N}^{n} Signal_{old}(i)$, where $N=5$ is the number of points the data was averaged over. The baseline noise due to the lower-frequency vibrations is subtracted by using the MATLAB function `polyfit' to fit a polynomial function to the signal, excluding the time interval for which plasma is present, and subtracting that polynomial approximation from the entire signal array.

Figs. 3(c) and 3(d) show the line-integrated electron density and phase shift measurements over multiple chords, where Chord 8 is 35 cm from the railgun nozzle and consecutive chords are placed at 6.35 cm intervals further along the axis of jet propagation. Due to the curvature of the spherical vacuum chamber and the choice of diagnostic ports, this chord arrangement has an average path length mismatch of 2 cm between consecutive chords, a cumulative path length mismatch of $\approx 14$ cm between Chord 8 and Chord 1, and an average path length mismatch of 20 cm between probe and reference path lengths. Limited space on the breadboards at the chamber forces optics placement which can add up to an additional 12 cm of path length discrepancy. In addition to density information, comparison of the density profiles between chords can also yield jet velocity and expansion data. Shot 30 had a velocity of 15 km/s and Shot 31 had a velocity of 26 km/s, where the velocities are calculated using the time delay between the leading edges of the jets, as shown in Figs. 3(c) and 3(d). The multiple peaks in each trace are likely due to electrical current ringing in the plasma gun source. Each ring produces a local density peak in the plasma jet.  These features will be investigated further in other work and do not directly affect the results of this paper.

\section{Summary}
In summary, an eight-chord, heterodyne, visible, fiber-optic interferometer achieves sub-fringe resolution even with path length discrepancies on the order of tens of centimeters between each probe-chord path length and the reference path length. The ability to have path length discrepancies in the system without signal degradation allows reduction in the complexity and cost of the interferometer by permitting the use of only one reference beam for all probe beams. The ability to have path length discrepancies also makes path length variations created by the spherical geometry of the vacuum chamber and limited space for optics a non-issue. The fiber-optic decoupling of the optics at the chamber from the rest of the interferometer greatly increases the flexibility of possible chord beam arrangements, and thus increases the range of accessible experiments. Finally, the sub-fringe resolution of the interferometer allows detection of line-integrated electron densities down to $10^{15} \mbox{ cm}^{-2}$ in addition to the $10 ^{17} - 10^{18} \mbox{ cm}^{-2}$ densities expected from the HyperV Mark I plasma railgun experiments.

\begin{acknowledgments}
The authors would like to acknowledge the contributions from their colleagues Andrew Case, Sarah Messer, and Samuel Brockington at HyperV Technologies, Jason Cassibry at the University of Alambama in Huntsville, and Thomas Awe, John Dunn, Colin Adams and Josh Davis at Los Alamos National Laboratory. This work was supported by the Office of Fusion Energy Sciences of the U.S. Dept. of Energy.
\end{acknowledgments}

\end{document}